# Theoretical Studies on Sodium Storage Mechanism in Hard Carbon Anodes of Sodium-Ion Batteries: Molecular Simulations Based on Machine Learning Force Fields

Zhaoming Wang[1,4], Guanghui Shi[1,4], Guanghui Wang[1,4], Man Wang[1,4], Xiao Wang[1], Feng, Ding[1,2,3,*]

1. Institute of Technology for Carbon Neutrality, Shenzhen Institute of Advanced Technology, Chinese Academy of Sciences, Shenzhen 518055, China
2. Suzhou Laboratory, Suzhou 215123, China
3. Faculty of Materials Science and Energy Engineering, Shenzhen University of Advanced Technology, Shenzhen 518055, China
4. Nano Science and Technology Institute, University of Science and Technology of China, Suzhou 518107, China
5. State Key Laboratory of Structural Analysis for Industrial Equipment and School of Physics, Dalian University of Technology, Dalian 116024 China

Email: xiao. f.ding@siat.ac.cn

**Abstract**

Sodium-ion batteries (SIBs) have garnered significant attention in recent years as a promising alternative to lithium-ion batteries (LIBs) due to their low cost, abundant sodium resources, and excellent cycling performance[1,2]. Hard carbon materials, characterized by their high specific capacity, outstanding cycling stability, and low cost, have emerged as potential candidates for SIB anodes[3,4]. However, the sodium storage mechanism in hard carbon anodes remains highly complex, especially in disordered structures, and is yet to be fully understood[5,6]. To address this, we employed relative machine learning force fields (MLFFs) in conjunction with multiscale simulation techniques to systematically investigate the sodium storage behavior in hard carbon[7,8]. By integrating simulations, this study provides a detailed exploration of sodium adsorption, intercalation, and filling mechanisms. High-precision, large-scale simulations reveal the dynamic behavior and distribution patterns of sodium ions in hard carbon. The findings not only deepen our understanding of sodium storage mechanisms in hard carbon anodes, but also offer a theoretical foundation for optimizing future SIB designs, while introducing novel simulation methodologies and technical frameworks to enhance battery performance[9-12].

## Introduction

Sodium-ion batteries (SIBs) have emerged as a promising alternative to lithium-ion batteries (LIBs) due to their abundant sodium resources, low cost, and exceptional cycling performance[1,13]. These advantages make SIBs particularly attractive for large-scale energy storage and electric vehicle applications, where their high energy density, long cycle life, and cost-effectiveness offer significant potential[14,15]. Among the various components of SIBs, the choice of anode material is critical. Hard carbon, with its high specific capacity, excellent cycling stability, and low cost, has become a leading candidate for SIB anodes[16,17]. However, despite its widespread application, the sodium storage mechanism in hard carbon remains highly complex. This complexity arises primarily from the disordered structure and high defect density of hard carbon, posing unresolved scientific questions about how sodium ions interplay with such materials at the microscopic level[18]. These structural features significantly influence the storage and diffusion behavior of sodium ions, but previous experimental studies, while providing preliminary insights, have yet to offer a clear understanding of how these microscopic structures impact sodium storage mechanisms[5,19].

The charge and discharge processes in SIBs exhibit distinct voltage plateaus, reflecting fundamental differences in sodium storage mechanisms at high and low voltage regions[20-23]. In the high-voltage region, sodium storage is primarily governed by adsorption and intercalation mechanisms. Sodium ions are initially adsorbed onto defect sites or the surface of hard carbon before intercalating into the graphene layers, resulting in minor voltage variations[24,25]. This behavior enables sodium ions stored in the low-voltage region to exhibit high cycling stability. Conversely, in the low-voltage region (close to 0 V), sodium storage is dominated by filling and intercalation mechanisms[26-29]. Sodium ions aggregate into clusters within the nanopores of hard carbon, leading to significant voltage variations. This clustering process in the high-voltage region often involves pronounced structural changes and stress accumulation, posing challenges to the performance and cycling stability of the battery. Despite progress, the understanding of sodium storage mechanisms in hard carbon remains incomplete, particularly regarding the interplay of different mechanisms and the precise origin of voltage plateaus[30-32].

Existing theoretical models have proposed several key mechanisms for sodium storage in hard carbon. Early studies focused primarily on adsorption, suggesting that sodium ions are adsorbed onto the surface or defect sites of hard carbon [33,34]. Other studies emphasized intercalation, where sodium ions are stabilized by van der Waals forces as they insert between graphene layers[35,36]. Additionally, a filling mechanism has been proposed, especially in low-voltage regions where high sodium concentrations lead to the formation of sodium clusters within nanopores[37]. However, these models often rely on simplified structural assumptions, neglecting the effects of disordered structures and high defect densities in hard carbon[38,39]. This limitation has left core questions unresolved, including the precise role of hard carbon's microstructure in governing sodium adsorption, intercalation, and filling mechanisms, as well as the interaction between these mechanisms. The lack of a unified and comprehensive understanding of sodium storage processes within hard carbon remains a significant challenge in the field[40,41].

Our study utilizes GPUMD software and machine learning force field (MLFF) simulation method to thoroughly analyze the sodium storage mechanisms in hard carbon[7,42]. Through high-precision multiscale simulations, we investigated how the microscopic structure of hard carbon affects sodium ion adsorption, intercalation, and filling behaviors, as well as the differences in storage mechanisms between low- and high-voltage regions[43,44]. These simulations not only provide a clearer understanding of sodium storage mechanisms in hard carbon but also reveal how its microstructure influences electrochemical performance, offering theoretical guidance for the optimization of hard carbon anode materials.

## 2. Results and Discussion

### 2.1 Construction and Validation of the Machine Learning Force Field Model

As illustrated in **Figure 1a**, the construction of the machine learning force field (MLFF) model involves a multi-step process of training dataset generation and iterative optimization. In this study, the initial training dataset was generated by combining first-principles molecular dynamics (AIMD) and perturbation methods. Using Vienna Ab initio Simulation (VASP) software, we constructed an initial dataset covering five distinct types of sodium-carbon structures: disordered carbon (hard carbon), layered carbon (e.g., graphene and graphite), spherical carbon materials (e.g., fullerenes), carbon nanotubes, and pure metallic sodium[46,47].

For disordered carbon structures, we simulated the effects of varying defect densities and porosities on sodium storage behaviors, focusing on sodium adsorption at defect sites, intercalation, and filling behaviors. In layered carbon materials, we analyzed the storage of sodium ions through intercalation between graphene layers and examined how interlayer spacing and defect levels influence the storage behavior[48-50]. Spherical carbon materials (e.g., fullerenes) were studied for their ability to provide diverse storage spaces, and simulations explored sodium intercalation on the surface and inside these structures[51]. For carbon nanotubes, the focus was on sodium storage both on the surface and inside the tubes, with particular attention to the effects of tube curvature and defect densities on intercalation behavior[52,53]. Finally, for benchmark, we calculated the interactions between sodium ions and pure metallic sodium to provide a reference for comparison[54].

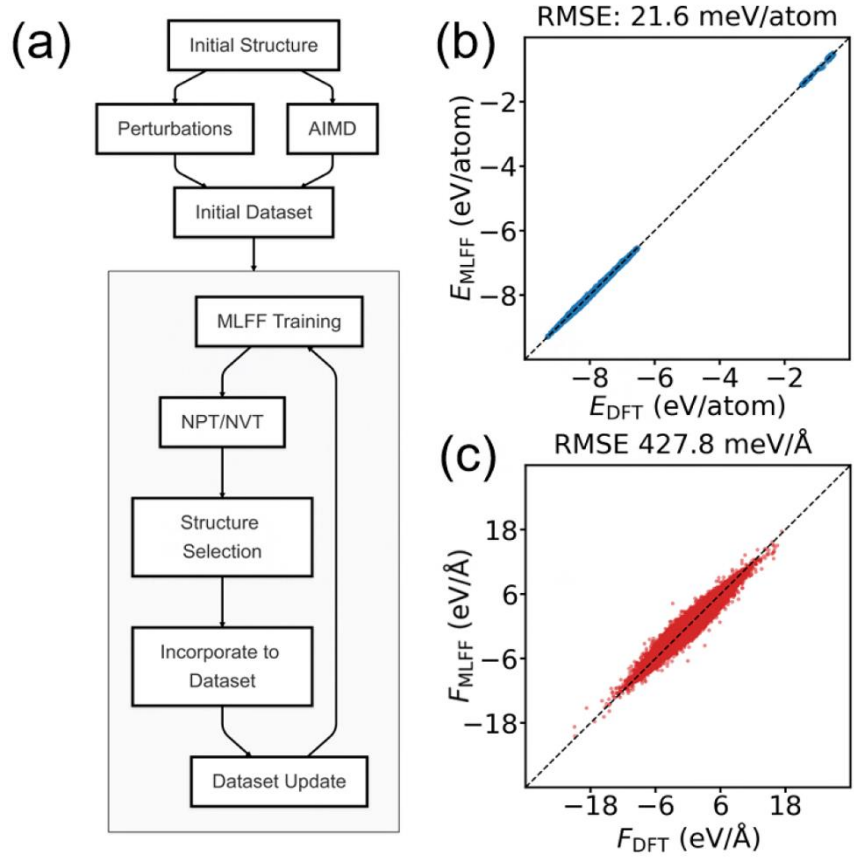

**Figure 1. The Training of the machine learning force field (MLFF).** (a) Workflow for training the MLFF; (b) Energy deviation for the test set; (c) Force deviation for the test set.

After constructing the initial dataset, the model was iteratively refined under NPT or NVT ensembles by evaluating its predictions and incorporating structures with remarkable errors or physical character back into the dataset. This process of dynamic selection and inclusion continued until the model predictions for energy and forces converged to a satisfactory accuracy.

As shown in **Figures 1b** and **1c**, the final MLFF model achieved excellent accuracy in the sodium-disordered carbon system, with a root mean square error (RMSE) of 0.022 eV/atom for energy predictions and 0.428 eV/Å for force predictions. These results demonstrate the reliability and precision of the MLFF model in capturing the complex interactions between sodium ions and carbon-based materials.

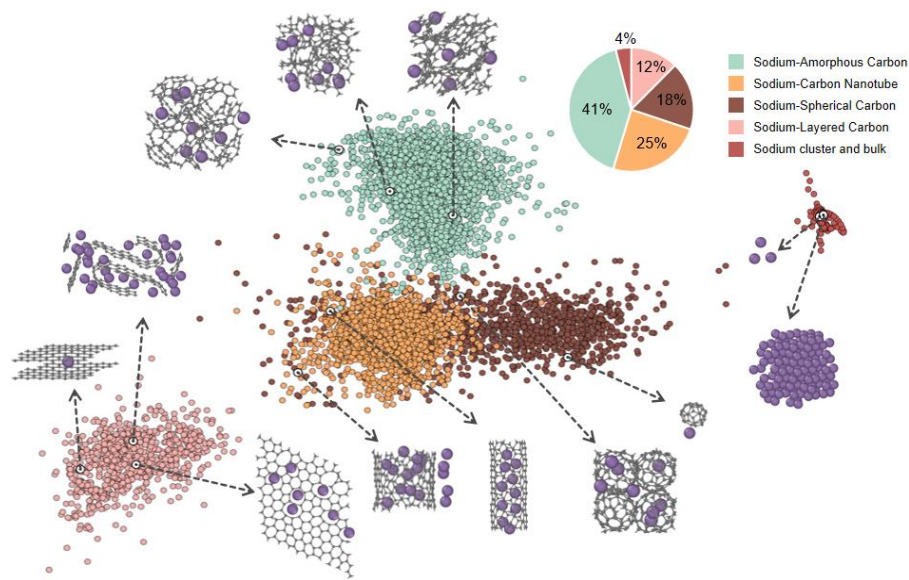

**Figure 2.** Visualization of the dataset with component distribution. The green region represents the amorphous carbon structure with a carbon-sodium content of 41% (2547 structures). The pink region indicates the layered carbon-sodium structure (12%, 763 structures), the yellow region represents the carbon nanotube-sodium structure (25%, 1507 structures), the brown region shows the spherical carbon-sodium structure (18%, 1094 structures), and the blue region represents pure metallic sodium (4%, 245 structures).

Later, we tested the accuracy of the as-trained MLFF. Table 1 presents a comparison of DFT-optimized (VASP optimized structure) and MLFF-optimized (MLFF optimized structure) results across different structures, including four randomly generated disordered carbon configurations and two graphene edge structures (zigzag edge and armchair edge) with sodium adsorption. In the first four disordered carbon configurations, sodium ions were randomly inserted into the disordered carbon framework and subsequently optimized using both DFT and MLFF methods. The results show a high level of agreement between MLFF and DFT optimizations in terms of total energy and atomic positions, with relative energy errors consistently below 0.6%. This demonstrates that the MLFF model can accurately exhibit the configuration, energy and force based on our MLFF, exhibiting strong predictive capability and reliability.

**Table 1.** Validation of the structures based on MLFF. From left to right, four rows represent randomly generated disordered carbon-sodium structures. From top to bottom: VASP-optimized structures and MLFF-optimized structures. The energy of each atom is displayed above each structure, and the error between VASP and MLFF optimization is shown below. The RMSE is 0.028 eV/atom.

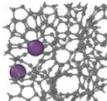

The last two columns in Table 1 compare sodium adsorption behavior on zigzag and armchair graphene edges. For these two graphene edge structures, the sodium ion positions and energy distributions optimized by DFT and MLFF exhibit remarkable consistency. For the zigzag edge, the energy calculated by DFT is 8.466 eV/atom, while MLFF predicts an energy of 8.445 eV/atom, with a relative error of only 0.25%. Similarly, for the armchair edge, the energy discrepancy is just 0.35%. These results further validate the accuracy of the as-trained MLFF model in describing sodium adsorption mechanisms on different graphene edge configurations, enabling the reliable characterization of sodium-carbon interactions in various microenvironments.

## 2.2 Construction of Hard Carbon Models and Sodium Ion Insertion Simulations

Following the development of the MLFF model, the next step involves constructing hard carbon models and simulating the sodium ion insertion process. Hard carbon, as a widely used anode material in sodium-ion batteries, exhibits significant impacts on sodium storage behavior due to its disordered structure and high defect density. Therefore, accurately reconstructing the microscopic structure of hard carbon is essential for simulating sodium storage mechanisms[55-59].

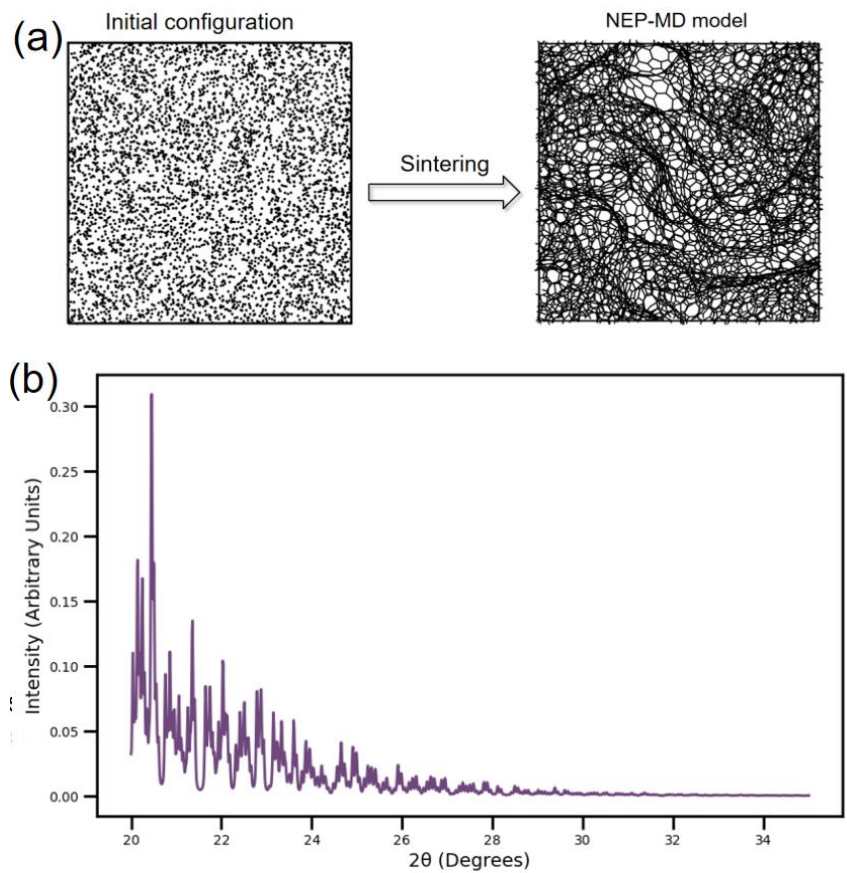

**Figure 3.** The hard carbon model using our MLFF. (a) The left side shows the initial structure, where carbon atoms are randomly inserted in the box. The right side displays the NEP-MD model, representing disordered structures generated by molecular dynamics (MD) simulations using MLFF. (b) The X-ray diffraction (XRD) spectrum of disordered carbon generated by the NEP-MD model. The horizontal axis represents the diffraction angle $2\theta$, and the vertical axis represents intensity.

The construction of the hard carbon model is aimed at not only replicating the disordered structural characteristics of the material but also enriching the environmental diversity of sodium storage sites[60-64]. The construction of the hard carbon model employed a well-established carbon potential function and a high-temperature annealing method to generate disordered hard carbon structures[73-75]. As illustrated in **Figure 3**, the initial configuration was generated by randomly inserting carbon atoms into a predefined spatial domain, followed by a high-temperature annealing process to simulate the interactions between carbon atoms[65,66]. This approach ultimately yielded a hard carbon model characterized by typical disorder and high porosity. The introduction of disordered structures enhances both the number and diversity of storage sites, creating a more realistic representation of the sodium storage environment[67-69]. This diversity provides a solid foundation for investigating the adsorption, intercalation, and aggregation behaviors of sodium ions under various microenvironments, thereby improving the fidelity of simulations and contributing to a deeper understanding of sodium storage mechanisms[70-72].

As illustrated in **Figure 3b**, the XRD pattern reveals the typical characteristics of hard carbon. A broad and diffuse (002) peak appears in the 20°–30° range, indicating the disordered nature of interlayer stacking. The absence of higher-order diffraction peaks suggests a lack of long-range order in the layered arrangement. Furthermore, the overall low diffraction intensity and the absence of sharp peaks confirm that hard carbon possesses a highly disordered, amorphous structure. These structural features, such as increased interlayer spacing and high disorder, make hard carbon well-suited for providing efficient pathways for ion storage and migration in sodium-ion batteries.

The sodium ion insertion process in hard carbon was simulated using the Monte Carlo method. Initially, spatial coordinates of sodium ions within the disordered hard carbon structure were generated. Sodium ions were then inserted randomly into the hard carbon, ensuring that they maintained a minimum distance from the nearest carbon atom to avoid direct overlap. Only coordinates meeting this criterion were accepted for further simulation.

After inserting the sodium ions, as shown in **Figure 4**, structural optimization and voltage calculations were performed using MD-NEP simulations implemented in Large-scale Atomic/Molecular Massively Parallel Simulator (LAMMPS) package. A gradual insertion approach was adopted, with 50 sodium ions inserted at each iteration, followed by structural optimization to minimize the system energy. To enhance computational efficiency, sodium positions were selected randomly from the coordination list, focusing on low-energy configurations. Through multiple iterations of insertion and optimization, stable sodium-inserted structures were obtained[76-84].

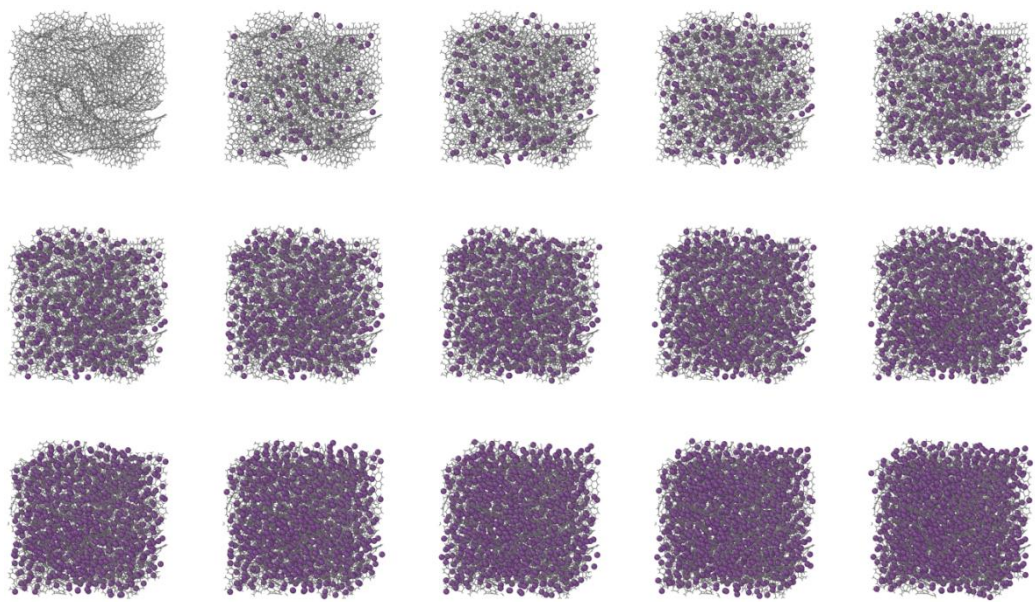

**Figure 4.** The Na-ion storage process, where 50 sodium atoms are randomly inserted in each iteration. A new image is generated after every two iterations.

**2.3 Interactions Between Sodium and Hard Carbon**

In this study, we conducted an in-depth analysis of the interactions between sodium ions and hard carbon, building upon and extending existing research[85]. By integrating experimental results with theoretical simulations, based on an analysis of the local coordination between sodium and carbon atoms, we further categorized sodium storage mechanisms into three types: adsorption, intercalation, and filling, which aligns with the conclusion of Surta et al.

Specifically, our analysis reveals that sodium adsorption primarily occurs at defect sites within the hard carbon, consistent with the surface adsorption sites described by Surta et al[85]. Intercalation mechanisms involve sodium ions being inserted into the voids between graphene layers, resembling the wrinkle sites reported in the literature. The filling mechanism, however, occurs in larger pore regions where sodium ions aggregate into clusters. Unlike the interlayer sites described by previous studies, our findings suggest that the hard carbon material used in this study may have larger pore structures, facilitating the formation of sodium clusters.

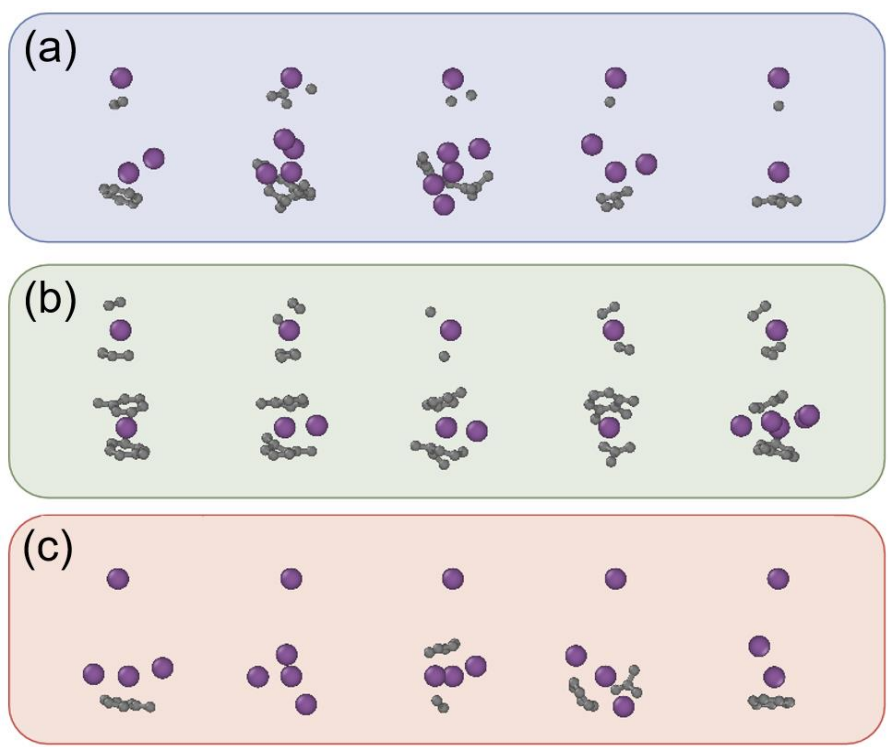

**Figure 5 The insertion site identification of Na-ion.** (a) adsorption sites; (b) intercalation site; (c) filling site. Five Na-ion insertion samples were randomly selected. In each subfigure, the upper panel shows the environment of sodium atoms within a cutoff radius of 2.7 Å, while the lower panel shows the environment within a cutoff radius of 3.5 Å, illustrating the overall coordination environment of the concentrated Na-ion.

Further we examined the microscopic distribution of sodium ions within hard carbon to elucidate their storage behavior. **Figure 5** illustrates sodium storage at different sites within the hard carbon. The results show that at low sodium concentrations, adsorption and intercalation dominate, with sodium ions preferentially occupying surface defects and interlayer voids[86]. At higher concentrations, filling becomes more prominent, with sodium ions aggregating into clusters within larger pores. However, we redefine sodium ions located on the cluster surface, which are in close proximity to carbon atoms (as indicated by the first peak in the radial distribution function), as adsorbed ions[87]. These findings are broadly consistent with the conclusions of Surta et al., but our study provides additional quantitative evidence, such as the number of sodium-carbon interactions within specific cutoff radii, to further validate the distribution of sodium ions across different sites[88-91].

## 2.4 Calculation of Formation Energy and Voltage Curves

The calculation of formation energy was and voltage curves plays a crucial role in understanding the sodium storage mechanisms in hard carbon anodes. Formation energy calculations provide insights into the thermodynamics of sodium ion insertion into hard carbon, revealing the stability and energetics of different storage behaviors. Voltage curves, on the other hand, directly reflect the electrochemical performance of the battery, capturing the processes of sodium ion insertion and extraction. As follows, we describe in detail the methodology for calculating formation energy and the corresponding voltage curves.

For the insertion and extraction processes of sodium ions in hard carbon, the voltage curve reflects the energy changes associated with sodium ion insertion and removal. The voltage can be calculated based on the variation in formation energy, expressed as:

$$V = -\frac{1}{e}\frac{E_f(x_2)-E_f(x_1)}{x_2-x_1} \tag{1}$$

where, e represents the elementary charge of an electron, $x_1$ and $x_2$ denote the number of sodium ions at the beginning and end of the insertion process, respectively, $E_f(x_1)$ and $E_f(x_2)$ are the formation energies corresponding to these two states. The voltage, expressed in volts (*V*), captures the energy changes during the insertion/extraction of sodium ions in hard carbon.

As shown in **Figure 6**, we calculated the voltage profile of sodium ions in hard carbon using the aforementioned methodology, with the results represented by the scatter points in the figure. Additionally, we analyzed the specific contribution of each sodium ion to the overall voltage. The solid line in the figure indicates the simulated half-cell voltage curve, while the scatter points represent the voltage contributions from individual sodium atoms. The colors of the scatter points indicate the classification of different storage sites as described in Section 2.3, including adsorption, intercalation, and filling. As shown in **Figure 6**, at the initial stages of sodium insertion, the formation energy decreased linearly with an increase in sodium content. As the sodium concentration continued to rise, the rate of energy reduction gradually slowed, eventually reaching a plateau.

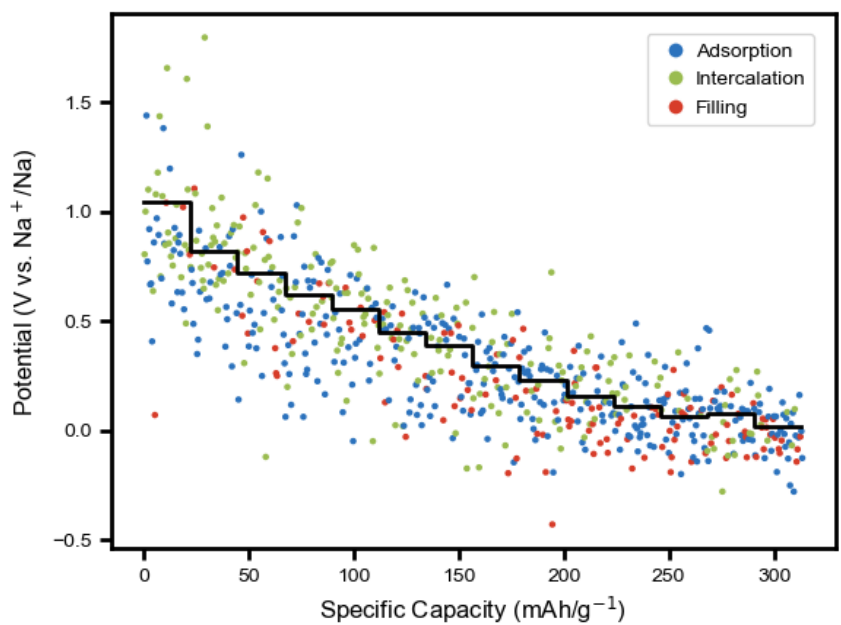

**Figure 6.** The voltage curve based on our hard model during half-cell charging, resulting in a step-like curve. The simulation is terminated when the voltage drops to 0. The three colors represent specific points under different voltages: blue for adsorption, green for intercalation, and red for filling.

The simulation results show that at high sodium concentrations, the voltage curve exhibits a distinct voltage plateau. In the low sodium concentration region, insertion primarily occurs through adsorption and intercalation, with minimal filling. Consequently, the voltage changes rapidly. In the high concentration region, filling becomes more prevalent, resulting in a plateau in the voltage curve. The height of the voltage plateau is directly related to the insertion energy of sodium ions, and the differences in plateau height can be explained by the variation in formation energy.

Due to the small size of the model (43.6 × 43.6 × 43.6 Å³) and the construction approach involving random carbon atom insertion followed by annealing, the model features a highly disordered and uniformly distributed carbon framework with small, densely packed pores. In this structure, sodium clusters formed in the pores tend to be relatively small, with their surface sodium atoms located close to the carbon atoms—approaching the first peak in the radial distribution function (RDF). Therefore, we defined these surface sodium atoms as adsorption states. This definition implies that, although adsorption is the dominant mechanism in the low voltage region, intercalation and filling mechanisms also contribute to the overall capacity, resulting in a rapid voltage change with capacity, indicative of the high energy gradient associated with adsorption and intercalation[92,93].

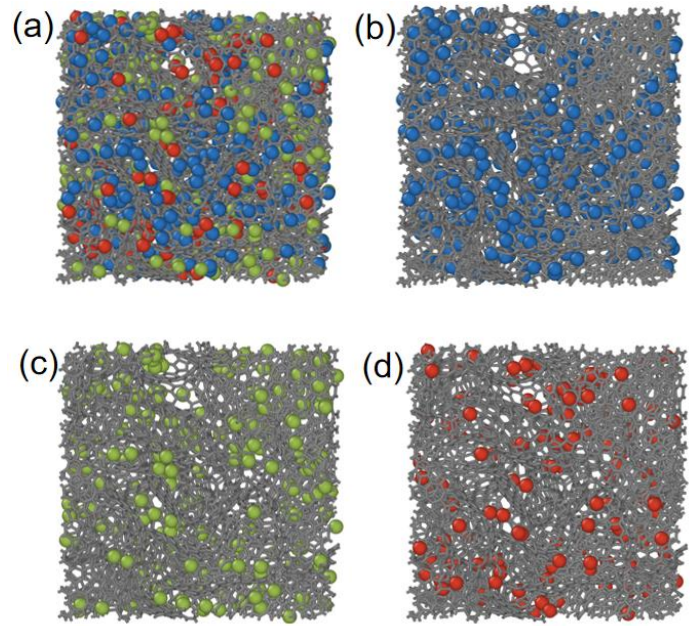

**Figure 7.** Verification of 700 sodium adsorption and intercalation sites under overcharged conditions, with twice the original capacity. (a) Shows the distribution of 700 sodium adsorption and intercalation sites;(c - d) Further illustrates these sites, with 340 atoms for adsorption, 223 for intercalation, and 137 for filling.

Furthermore, to explore the performance of sodium-ion batteries under different conditions, we also simulated the overcharged state of sodium ions in hard carbon. As illustrated in **Figure 7**, we further validated the adsorption and intercalation sites of sodium ions under overcharged conditions. The figure presents the distribution of 700 sodium ions at adsorption, intercalation, and filling sites under overcharging, encompassing all three storage mechanisms. In **Figure 7a**, the positions of all 700 sodium ions are shown; and **Figure 7b-d** depict the three distinct storage modes, where blue, green and red represents adsorption, intercalation, and filling, respectively. Specifically, 340 atoms were identified as being in adsorption states, 223 atoms as intercalation, and 137 atoms as filling. Our simulations indicate that during overcharging, sodium insertion not only affects the shape of the voltage curve but also induces structural changes in the hard carbon. During this process, the voltage continues to drop and eventually stabilizes, suggesting that excessive sodium insertion may compromise the structural stability of hard carbon, as detailed in the Supplementary Information.

## 4. Conclusions

This study comprehensively investigated the sodium storage mechanisms in hard carbon anodes using a multiscale simulation approach based on the MLFF. We successfully build the hard carbon model and recognized the three different inserting sites of Na ions during the charging process, i. e. absorption, intercalation and filling, and thus confirms the three corresponding insertion stage. Further simulations of the overcharged state revealed that excessive sodium insertion may cause structural changes in the hard carbon material, ultimately affecting the performance and stability of the battery. This study provides comprehensive theoretical support for understanding the sodium storage mechanisms in hard carbon, as well as new simulation methodologies and technical tools for the design and optimization of sodium-ion batteries. By employing a multiscale simulation approach, we gained in-depth insights into the storage processes of sodium ions in hard carbon, offering valuable reference points for the development and application of future sodium-ion batteries.

**Author contributions**

F.D. and X.W. conceived and supervised the project. Z.W. conducted the literature search, selected relevant studies and wrote all the sections of the article. G.S., G.W.,

M.W. discussed the technical details, F.D. and X.W. reviewed and edited the manuscript. All authors reviewed and approved the final manuscript.

**Data availability**

Data will be made available on request.

**Acknowledgments**

The authors acknowledge the support of the National Natural Science Foundation of China (22333005, 52372054), the Youth Innovation Promotion Association CAS, the High-Talent Grant from Shenzhen Institute of Advanced Technology (SIAT-SE3G0991010, 2023) and the startup grant from Shenzhen University of Advanced Technology.

**Disclosure statement**

No potential conflict of interest was reported by the authors.